
\magnification 1200
\vsize=21.0cm
\hsize=14.1cm
\baselineskip=6truemm

\def\ubar{\overline{u}}
\def\cbar{\overline{c}}
\def\tbar{\overline{t}}
\def\dbar{\overline{d}}
\def\sbar{\overline{s}}
\def\bbar{\overline{b}}
\def\fbar{\overline{f}}

\def\Abar{\overline{A}}
\def\Kbar{\overline{K}^0}
\def\Bbar{\overline{B}^0}

\def\bd{B^0}

\def\Dbar{\overline{D}^0}

\def\Gammabar{\overline{\Gamma}}
\def\to{\rightarrow}

\line{\hfil TECHNION-PH-95-25}
\line{\hfil August, 1995}
\null\vskip 1.5truecm

\centerline{\bf A VARIETY OF CP VIOLATING B DECAYS}
\vskip 10mm
\centerline{\it Michael Gronau}
\centerline{\it Department of Physics,
Technion -- Israel Institute of Technology}
\centerline{\it 32000 Haifa, Israel}

\vskip 2.0cm

\magnification 1200
\baselineskip=6truemm
\noindent
\centerline{\bf Abstract}
\vskip 3mm
\noindent
A variety of CP violating asymmetries in $B^0$ decays
is predicted in the Standard Model in terms of
CKM (Cabibbo-Kobayashi-Maskawa) complex phases. These phases can also be
determined from rate measurements of certain $B^+$ and $B^-$ decays. We focus
on the following processes: $B^0\to\psi K_S,~\pi^+\pi^-,~\pi^0\pi^0,
{}~\rho^{\pm}\pi^{\mp},~B_s\to D^+_s K^-,~\pi^0\eta,
{}~B^+\to D^0(\Dbar)K^+,~\pi^+\pi^0,~\pi^0 K^+,~
\pi^+ K^0,~\eta K^+$ and their charge-conjugates.
Complications due to gluonic-penguin and
electroweak-penguin amplitudes are dealt with and are resolved. The importance
of final state interaction phases in charged $B$ decay asymmetries
is demonstrated in $B^+\to\chi_{c0}\pi^+,~~\chi_{c0}\to\pi^+\pi^-$ through the
effect of the $\chi_{c0}$ width.

\vskip 5.5cm
\centerline{\it Invited talk presented at Beauty 95}
\centerline{\it Third International Workshop on B-Physics at Hadron Machines}
\centerline {\it Oxford, July 10-14, 1995}

\vfill \eject

\baselineskip=6truemm
\noindent
{\bf 1. Introduction}
\vskip 3mm

The Standard Model [1] accounts for the observed CP
violation in the neutral $K$ meson mixing [2] through a phase in
the Cabibbo-Kobayashi-Maskawa (CKM) matrix [3].
The $B$ meson system provides a wide variety of
independent CP asymmetry measurements, which can test the hypothesis that this
phase is the only source of CP violation in the presently known fundamental
interactions among elementary particles. The purpose of this talk is to
demonstrate some of the most promising ways of carrying out such tests. Since
new ideas, theoretical as well as experimental, are constantly being developed
in this field, I decided to combine in this talk some early and already
"classical" examples with very recent suggestions. For earlier reviews of
this subject, describing other processes and containing
a more complete list of references, see ref. [4].

Section 2 introduces the CKM matrix and summarizes the available
information on the magnitudes and phases of its elements. The subsequent two
sections deal with neutral and with charged $B$ decays.
Section 3 describes CP violation which occurs when mixed neutral
$B$ mesons decay to states which are common decay products of $B^0$ and
$\Bbar$.
Direct CP violation in charged $B$ decays is the subject of Section 4, while
Section  5 concludes.

\vskip 0.5truecm
\noindent
{\bf 2. CP violation in the standard model}
\vskip 3mm

In the standard model of three families of quarks and leptons the
$SU(3)_C\times SU(2)_L\times U(1)_Y$  gauge group is spontaneously
broken by the vacuum expectation value of a single scalar Higgs doublet.
CP violation occurs in the interactions of the three families of the
left-handed quarks with the charged gauge boson:
$$
-\cal L =
\left(
\matrix{\ubar&\cbar&\tbar\cr}\right)
\left(\matrix{m_u&~&~\cr ~&m_c&~\cr ~&~&m_t\cr}\right)
\left(\matrix{u\cr c\cr t\cr}\right)
+ \left(
\matrix{\dbar&\sbar&\bbar\cr}\right)
\left(\matrix{m_d&~&~\cr ~&m_s&~\cr ~&~&m_b\cr}\right)
\left(\matrix{d\cr s\cr b\cr}\right)
$$
$$
+{g\over\sqrt{2}}
\left(
\matrix{\ubar&\cbar&\tbar\cr}\right)_L \gamma^{\mu} V
\left(\matrix{d\cr s\cr b\cr}\right)_L W_{\mu}^+ +...
\eqno(1)
$$
The unitary CKM mixing matrix $V$ is defined [5] in terms of three
Euler-like family-mixing angles , $\theta_{ij}$, and a phase
$\gamma\equiv\delta_{13}$ which is the mere origin of CP violation.
The measured values of the three mixing angles, which have a hierarchical
pattern, are [6]:
$$
\sin\theta_{12}\equiv\vert V_{us}\vert=0.220\pm 0.002~,
$$
$$
\sin\theta_{23}\equiv\vert V_{cb}\vert=0.039\pm
0.005~,
$$
$$
\sin\theta_{13}\equiv\vert V_{ub}\vert=0.0035\pm 0.0015~.\eqno(2)
$$
The only information about a nonzero value of $\gamma\equiv{\rm Arg}(V^*_{ub})$
comes from CP violation in the $K^0-\Kbar$ system, which provides very loose
bounds on this phase (see eq. (4) below).

Unitarity of $V$ implies triangle relations such as
$$
V_{ud}V^*_{ub}+V_{cd}V^*_{cb}+V_{td}V^*_{tb}=0~,\eqno(3)
$$
which is shown in Fig. 1.
The three angles of the unitarity triangle, $\alpha,~\beta$ and $\gamma$
are rather badly known at present. Current
constarints, which depend on uncertainties in $K$- and $B$-meson hadronic
parameters, can be summarized by the following ranges [6]:
$$
10^0\leq\alpha\leq 150^0,~~~5^0\leq\beta\leq 45^0,~~~20^0\leq\gamma\leq 165^0~.
\eqno(4)
$$
As we will show in the following two sections, measurement of certain CP
asymmetries in $B$ decays are directly related to these three angles in a
manner
which is free of hadronic uncertainties.

We note in passing that in the standard phase convention [5] only two of the
CKM elements,
$V_{ub}$ with phase $-\gamma$, and $V_{td}$ with phase $-\beta$, carry a
complex phase. All other matrix elements are real to a good
approximation.

\vskip 5mm
\noindent
{\bf 3. CP violation in decays of mixed $B^0-\Bbar$}
\vskip 3mm
\noindent
{\it 3.1 Decays to CP eigenstates dominated by a single CKM phase}~[7]
\vskip 3mm

Consider the time-evolution of a state which is identified ("tagged") as
a $B^0$ at time $t=0$:
$$
t=0:~~~~|B^0\rangle ={e^{-i\phi_M}\over\sqrt{2}}(|B_L\rangle+|B_H\rangle)~.
\eqno(5)
$$
$B_{L,H}$ are the "light" and "heavy" mass-eigenstates and $\phi_M$ is the
phase of the $B^0-\Bbar$ mixing parameter [4], $\phi_M=\beta$ for $B^0$
and $\phi_M=0$ for $B_s$.
The $B_{L,H}$ state-evolutions are given by their masses and
by their approximately equal decay width $\Gamma$: $|B_{L,H}(t=0)\rangle \to
|B_{L,H}(t)\rangle= \exp[-i(m_{L,H}-{i\over 2}\Gamma)t]|B_{L,H}(t=0)\rangle$.
Thus, the $B^0$ oscillates into a mixture of $B^0$ and $\Bbar$:
$$
t:~~~~~~|B^0(t)\rangle=e^{-i\overline{m}t}e^{-{\Gamma\over 2}t}[\cos({\Delta m
t\over 2})|B^0\rangle+ie^{-2i\phi_M}\sin({\Delta mt\over 2})|\Bbar\rangle]~,
\eqno(6)
$$
where $\overline{m}\equiv(m_H+m_L)/2,~\Delta m\equiv m_H-m_L$.

Let us consider decays into states $|f\rangle$ which are eigenstates of CP,~
$CP|f\rangle=\xi|f\rangle$ with eigenvalue $\xi=\pm 1$, and let us assume that
a single weak amplitude (or rather a single weak phase) dominates the
decay process. Both $B^0$ and $\Bbar$ decay to the state $f$, with
amplitudes $A=|A|\exp(i\phi_f)\exp(i\delta_f)$ and $\Abar=\xi|A|\exp(-i
\phi_f)\exp(i\delta_f)$, respectively. $\phi_f$ and $\delta_f$ are the weak and
strong phases, respectively. The former changes sign under charge-conjugation,
whereas the latter remains the same. The time-dependent decay rate
of an initial $B^0$ is
$$
\Gamma(t)\equiv\Gamma(B^0(t)\to f)=e^{-\Gamma t}|A|^2[1+\xi\sin2(\beta+\phi_f)
\sin(\Delta mt)]~,\eqno(7)
$$
and the corresponding rate for an initial $\Bbar$ is
$$
\Gammabar(t)\equiv\Gamma(\Bbar (t)\to f)=e^{-\Gamma
t}|A|^2[1-\xi\sin2(\beta+\phi_f)
\sin(\Delta mt)]~.
\eqno(8)
$$
The time-dependent CP asymmetry is then given by
$$
Asym.(t)\equiv {\Gamma(t)-\Gammabar(t)\over \Gamma(t)+\Gammabar(t)}=
\xi\sin2(\phi_M+\phi_f)\sin(\Delta mt)~,\eqno(9)
$$
while the time-integrated asymmetry is
$$
Asym.= \xi\big({\Delta m/\Gamma\over 1+(\Delta m/\Gamma)^2}\big )\sin2(\phi_M+
\phi_f)~.
\eqno(10)
$$
That is, in this case {\it the CP asymmetry measures a CKM phase with no
hadronic uncertainty}. The integrated asymmetry in $\bd$ decays may be as large
as $(\Delta m/\Gamma)/[1+(\Delta m/\Gamma)^2]=0.47$.

The best example of decays to CP-eigenstates is the well-known gold-plated
case of $\bd\to \psi K_S$, for which a branching ratio of about
$5\times 10^{-4}$ has already been measured [8]. In this case
$\phi_M=\beta,~\phi_f={\rm Arg}(V^*_{cb}V_{cs})=0,~\xi=-1$. Another
case is $\bd\to\pi^+\pi^-$, for which a combined branching ratio
$B(\bd\to\pi^+\pi^-~{\rm and}~K^+\pi^-)=(1.8\pm 0.6)\times 10^{-5}$ has been
measured [9], with a likely solution in which the two modes have about equal
branching ratios. In this case $\phi_f={\rm
Arg}(V^*_{ub}V_{ud})= \gamma,~\xi=1$. Consequently one has in these
two cases
$$
Asym.(\bd\to\psi K_S;t)=-\sin2\beta\sin(\Delta mt)~,
$$
$$
Asym.(\bd\to\pi^+\pi^-;t)=-\sin2\alpha\sin(\Delta
mt)~.\eqno(11)
$$
In the case of decay to two pions the asymmetry obtains, however, corrections
from a second (penguin) CKM phase. This problem will be discussed below.

\vskip 5mm
\noindent
{\it 3.2 Decays to non-CP eigenstates}~[10][11]
\vskip 3mm
\noindent

Angles of the unitarity triangle can also be determined from neutral B decays
to states $f$ which are not eigenstates of CP. This is feasible when
both a $B^0$ and a $\Bbar$ can decay to a final state which appears in only one
partial wave, provided that a single CKM phase dominates each of the
corresponding decay amplitudes.

The time-dependent rates for states which are $B^0$ or $\Bbar$ at $t=0$ and
decay at time $t$ to a state $f$ or its charge-conjugate $\fbar$ are given
by:
$$
\Gamma_f(t)=e^{-\Gamma t}[|A|^2\cos^2({\Delta mt\over 2})+|\Abar|^2\sin^2
({\Delta mt\over 2})+|A\Abar|\sin(\Delta\delta+\Delta\phi_f+2\phi_M)\sin(\Delta
mt)]~,
$$
$$\Gammabar_f(t)=e^{-\Gamma t}[|\Abar|^2\cos^2({\Delta mt\over
2})+|A|^2\sin^2 ({\Delta mt\over
2})-|A\Abar|\sin(\Delta\delta+\Delta\phi_f+2\phi_M)\sin(\Delta mt)]~,
$$
$$
\Gamma_{\fbar}(t)=e^{-\Gamma t}[|\Abar|^2\cos^2({\Delta mt\over
2})+|A|^2\sin^2 ({\Delta mt\over
2})-|A\Abar|\sin(\Delta\delta-\Delta\phi_f-2\phi_M)\sin(\Delta mt)]~,
$$
$$
\Gammabar_{\fbar}(t)=e^{-\Gamma t}[|A|^2\cos^2({\Delta mt\over
2})+|\Abar|^2\sin^2
({\Delta mt\over 2})+|A\Abar|\sin(\Delta\delta-\Delta\phi_f-2\phi_M)\sin(\Delta
mt)]~.\eqno(12)
$$
Here $\Delta\delta_f,~(\Delta\phi_f)$ is the difference between the strong
(weak) phases of $A$ and $\Abar$, the decay amplitudes of $B^0$ and $\Bbar$
to the common state $f$. The four rates depend on four
unknown quantities, $|A|,~|\Abar|,~\sin(\Delta\delta_f+\Delta\phi_f+2\phi_M),~
\sin(\Delta\delta_f-\Delta\phi_f-2\phi_M)$. Measurement of the rates allows a
determination of the weak CKM phase $\Delta\phi_f+2\phi_M$ apart from a
two-fold ambiguity.

There are at least two interesting examples to which this method has been
applied [11].
In the first case, $\bd\to\rho^+\pi^-$, one must neglect a second contribution
of a penguin amplitude, a problem which will be addressed in the following two
subsections. Assuming for a moment that tree diagrams
dominate $A$ and $\Abar$, one can measure in this manner the angle $\alpha$. A
second case, which may be used to measure $\gamma$, is $B_s\to D^+_s K^-$.

\vskip 5mm
\noindent
{\it 3.3 Corrections from penguin amplitudes}~[12]
\vskip 3mm

In a wide variety of decay processes, such as in $\bd\to\pi^+\pi^-$,
there exists a second amplitude due to a ``penguin" diagram in addition to
the usual ``tree" diagram. The two diagrams cary different CKM phases and in
general may have different strong phases.
In such a case CP is violated in the direct decay of
a $B^0$. Then in decays to CP-eigenstates one has $|A|\not=|\Abar|$, and the
asymmetry acquires a time-dependent cosine term in addition to the sine term:
$$
Asym.(t)=
{(1-|\Abar/A|^2)\cos(\Delta mt)-2{\rm
Im}(e^{-2i\phi_M}\Abar/A)\sin(\Delta mt)\over 1+|\Abar/A|^2}~.
\eqno(13)
$$

Observation of an extra $\cos(\Delta mt)$ term implies direct
CP violation and would invalidate the above method of measuring a weak phase.
The opposite is not true, however, since the absence of a cosine term
does not guarantee that the coefficient of the sine term is
given in terms of a CKM phase. The coefficient of $\cos(\Delta mt)$
is proportional to $\sin(\Delta\delta)$, where $\Delta\delta$ is the
final-state
phase-difference between the tree and penguin amplitudes. On the other hand,
the coefficient
of $\sin(\Delta mt)$ obtains a correction proportional to
$\cos(\Delta\delta)$. If $\Delta\delta$ were small this correction might
be large, in spite of the fact that the $\cos(\Delta mt)$ term were too small
to be observed. Assuming, for instance $\Delta\delta=0$ in $\bd\to\pi^+\pi^-$,
where the penguin-to-tree ratio of amplitudes may be as large as about 0.2, the
coefficient of $\sin(\Delta mt)$ may be as large as 0.4 for
$\sin(2\alpha)=0$~[13] (for which no asymmetry is expected in the absence of
penguins). On the other hand, the decay $\bd\to \psi K_S$ remains
a pure case, since in this case the penguin-to-tree ratio of amplitudes with
unequal weak phases is extremely small.

\vskip 5mm
\noindent
{\it 3.4 Removing penguin corrections in $\bd\to\pi^+\pi^-$}~[14]
\vskip 3mm

It is possible to disentangle the penguin contribution in $\bd\to\pi^+\pi^-$
from the tree-dominating asymmetry by measuring also the rates of
$B^+\to\pi^+\pi^0$ and $\bd\to\pi^0\pi^0$. No time-dependence is required
for these processes. The method is based on the
observation that the two weak operators contributing to these decays have
different isospin properties. Whereas the tree
operator is a mixture of $\Delta I=1/2$ and $\Delta
I=3/2$, the gluonic penguin operator is pure $\Delta I=1/2$.

The  physical amplitudes of $B\to \pi^+\pi^-,
\pi^0\pi^0, \pi^+\pi^0$  can be decomposed into their isospin components
$$
{1\over\sqrt{2}}A^{+-}=A_2-A_0~,~~~A^{00}=2A_2+A_0~,~~~A^{+0}=3A_2~,
\eqno(14)
$$
where $A_0$ and $A_2$ correspond to $\pi\pi$ states with $I=0$ and $I=2$,
respectively. This yields a complex triangle relation
$$
{1\over\sqrt{2}} A^{+-} + A^{00} = A^{+0}~,\eqno(15)
$$
and a similar triangle relation for the charge-conjugated processes
$$
{1\over\sqrt{2}}{\overline{A}}^{+-}
+ {\overline{A}}^{00} = {\overline{A}}^{-0}~.\eqno(16)
$$

Applying a simple phase rotation, $\tilde{A}=\exp(2i\gamma)\overline{A}$
($\gamma$ being the phase of the tree amplitude), the two triangles (15)(16)
are
described in Fig. 2, sharing a common base, $|A^{+0}|=|{\tilde{A}}^{-0}|$ (at
this point we neglect the
electroweak penguin contributions $P_{EW},~\tilde{P}_{EW}$). The angle $\theta$
between the sides of the two triangles corresponding to $A^{+-}$ and
${\tilde{A}}^{+-}$ and the ratio of the lengths of these sides determine, for
a given $\alpha$, the coefficient of the $\sin(\Delta mt)$ term in
$\bd(t)\to\pi^+\pi^-$:
$$
{\rm coefficient~of}~\sin(\Delta mt)={\vert\tilde{A}^{+-}\vert\over\vert
A^{+-}\vert}\sin(2\alpha+\theta)~.\eqno(17)
$$
Thus, the time-dependence of $\bd(t)\to\pi^+\pi^-$, and the integrated decay
rates involving neutral pions, determine $\alpha$. There exists a two-old
ambiguity in determining $\alpha$ due to the fact that one of the two triangles
in Fig. 2 may be turned up-side-down

Recently it was noted [15] that electroweak penguin contributions could
spoil this method, since these amplitudes are not pure $\Delta I=1/2$.
A closer look at the effect shows that, in fact, the uncertainty in determining
$\alpha$ remains very small [16]. The effect
is shown in Fig. 2, where the terms $P_{EW}$ and $\tilde{P}_{EW}$ represent the
electroweak penguin contributions to $B$ and $\overline{B}$ decays,
respectively.
As a result of these terms the two triangles, for $B$ and $\overline{B}$
decays,
do not share a common base. CP is violated also in $B^+\to\pi^+\pi^0$. This
introduces a small uncertainly in determining $\theta$ and subsequently
$\alpha$.
Model-dependent calculations [15] and more general order-of-magnitude
arguments [16] show the following hierarchy among the tree ($T$), gluonic
penguin ($P$) and electroweak penguin ($P_{EW}$) contributions in $B\to\pi\pi$
$$
T:P:P_{EW}\sim 1:\lambda:\lambda^2~,~~~\lambda=0.2~.\eqno(18)
$$
Therefore, the uncertainty in $\alpha$ from electroweak penguin amplitudes
is at most a few degrees:
$$
\Delta\alpha={1\over 2}\Delta\theta\leq\lambda^2~.\eqno(19)
$$
A similar isospin analysis can be carried out for $B\to\rho\pi$ [17].
To resolve certain ambiguities in $\alpha$, a full Dalitz plot analysis must
be made for the three pion final states.

\vskip 0.5truecm
\noindent
{\bf 4. CP violation in charged $B$ decays}
\vskip 3mm

\noindent
{\it 4.1 The problem}
\vskip 3mm

The simplest manifestation of $CP$ violation, which requires neither tagging
nor a time-dependent measurement, is finding different partial decay
widths for a particle and its antiparticle into corresponding decay modes.
Consider a general decay $B^+\to f$ and its charge-conjugate process
$B^-\to\fbar$. In order that these two proceses have different rates, two
amplitudes ($A_1, A_2$) must contribute, with different CKM phases ($\phi_1 \ne
\phi_2$) and different final state interaction phases ($\delta_1\ne\delta_2$):
$$
A(B^+\to f)~=~\vert A_1\vert  e^{i\phi_1}e^{i\delta_1}~+ ~\vert
A_2\vert e^{i\phi_2}e^{i\delta_2}~,
$$
$$
{}~~~~~\Abar (B^-\to \fbar)~=~\vert A_1\vert  e^{-i\phi_1}e^{i\delta_1}~+
{}~\vert
A_2\vert e^{-i\phi_2}e^{i\delta_2}~,
$$
$$
{}~~~~~~~~~\vert A \vert^2-\vert\Abar\vert^2=2\vert A_1
A_2\vert\sin(\delta_1-\delta_2)\sin(\phi_1-\phi_2)~. \eqno(20)
$$
For $|A_2|^2\ll |A_1|^2$ one finds a CP asymmetry
$$
Asym.={\vert A \vert^2-\vert\Abar\vert^2\over \vert A
\vert^2+\vert\Abar\vert^2}
\approx 2{|A_2|\over |A_1|}\sin(\delta_1-\delta_2)\sin(\phi_1-\phi_2)~.
\eqno(21)
$$

The theoretical difficulty of relating an asymmetry in charged $B$ decays to a
pure CKM phase, ($\phi_1-\phi_2$), follows from having two unknowns in the
problem: The
ratio of amplitudes, $\vert A_2/A_1\vert$, and the final state phase
difference,
$\delta_2-\delta_1$. Both quantities involve quite large theorertical
uncertainties. In this section I will discuss three kinds of
solutions to this problem:
\item{a.} Measure $A_1$ and $A_2$ independently ({\it 4.2}).
\item{b.} Relate $A_1$ and $A_2$ by symmetry to other directly measurable
amplitudes ({\it 4.3}).
\item{c.} Control $\delta_2-\delta_1$ as much as possible ({\it 4.4}).

\vskip 5mm
\noindent
{\it 4.2 Measuring $\gamma$ in $B^{\pm}\to D^0 K^{\pm}$}~[18]
\vskip 3mm

The decays
$B^{\pm}\to D^0_1(D^0_2) K^{\pm}$ and a few other processes of this
type provide a unique case, in which one can measure separately the
magnitudes of the two contributing amplitudes, and thereby determine the CKM
phase  $\gamma$.

$D^0_1(D^0_2)=(D^0+(-)\Dbar)/\sqrt{2}$ is a CP-even (odd)
state, which is  identified by its CP-even (odd) decay products. For
instance, the states $K_S\pi^0,~K_S\rho^0,~K_S \omega,~K_S \phi$ identify a
$D^0_2$, while $\pi^+\pi^-,~ K^+K^-$ represent a $D^0_1$.
For $D^0_1=(D^0+\Dbar)/\sqrt{2}$ we have
$$
\eqalign{
\sqrt{2}A(B^+\to D^0_1 K^+)~=~A(B^+\to
D^0 K^+)~+~A(B^+\to \Dbar K^+),\cr \sqrt{2}A(B^-\to D^0_1 K^-)~=~A(B^-\to \Dbar
K^-)~+~A(B^-\to D^0 K^-).\cr}\eqno(22)
$$
$D^0$ and $\Dbar$, states of specific flavor, are identified by the charge
of the decay lepton or kaon. The amplitudes of $B^+\to
D^0 K^+$ and $B^+\to \Dbar K^+$ are shown schematically in Fig. 3. Their CKM
factors,
$V^*_{ub}V^{~}_{cs}$ and $V^*_{cb}V^{~}_{us}$, are of comparable
magnitude. Their weak phases are $\gamma$ and zero, respectively. The two
relations (22) are described by the two triangles in Fig. 4 representing the
$B^+$ and $B^-$ decay amplitudes.

$\gamma$ is determined with a two-fold ambiguity by the
rates of the above six proccesses, two pairs of which are equal. Note that
$\gamma$ can be measured also when $\delta_1-\delta_2=0$, in which case no
asymmetry is observed and the two triangles in Fig. 4 must be drawn
up-side-down with respect to each other.
The feasibility of observing a CP asymmetry in $B^+\to D^0_{1(2)} K^+$ and the
precision of measuring $\gamma$ depend
on the branching ratios of the three related decay processes, and on the values
of the weak and strong phases. The decay $B^+\to D^0 K^+$ is expected to be
color-suppressed in addition to being CKM suppressed. Using a value
of $5\times 10^{-6}$ for the branching ratio of this process, the feasibility
for observing a CP asymmetry in $B^+\to
D^0_{1(2)} K^+$ was studied [19] as function of $\gamma$ and
$\delta_2-\delta_1$, for a (symmetric) $e^+e^-\to\Upsilon(4S)$ $B$-factory with
an integrated luminosity of $20 fb^{-1}$. The discovery region was found
to cover a significant part of the ($\gamma, \delta_2-\delta_1$) plane. For
small final state phase differences this experiment is sensitive mainly to
values
of $\gamma$ around $90^0$. Large values of $\delta_2-\delta_1$ allow a
useful measurement of $\gamma$ in the range $50^0\leq\gamma\leq 130^0$.
This method was generalized to
quasi-two-body decays  $B\to DK_i\to D K\pi$, where $K_i$ are excited kaon
resonance states with masses around 1400 MeV [20]. These resonances
give rise to large calculable final state phases and thus may enhance the
asymmetry.

\vskip 5mm
\noindent
{\it 4.3 Using SU(3) to determine $\gamma$ from $B^+\to\pi\pi,~\pi K,~\eta
K$}~[21][22]
\vskip 3mm

Flavor SU(3) symmetry can be used to obtain relations among a variety of $B$
decays [23]. Recently we have applied this symmetry and linearly broken SU(3)
to two body decays [24]. We neglected small
annihilation-like contributions which are expected to be suppressed by
$f_B/m_B$. This suppression can be tested, for instance, by pushing upper
limits on $B^0\to K^+ K^-$ down to a level of $10^{-7}$.
In the following discussion we use quark diagrams as a particularly useful
representation of SU(3) amplitudes.

Consider the decay $B^+\to \pi^0 K^+$ to which the two diagrams of Fig. 5
contribute. These diagrams represent a strong penguin and a tree amplitude,
and are related by symmetry to measurable amplitudes of other decay processes.
The penguin amplitude is related by isospin (exchanging $u\ubar$ by $d\dbar$)
to the amplitude of $B^+\to \pi^+K^0$. The annihilation contribution to the
latter process is neglected. The tree amplitude is related by SU(3) (exchanging
$\sbar$ by $\dbar$) to the amplitude of $B^+\to\pi^+\pi^0$, which receives
no strong
penguin contribution. SU(3) breaking can be introduced into this
relation by assuming factorization of tree amplitudes. Thus the tree amplitude
is given by $(f_K/f_{\pi})|V_{us}/V_{ud}|A(\pi^+\pi^0)$. One therefore
obtains a simple triangle relation between the three $B^+$ decay amplitudes:
$$
A(\pi^0 K^+)-{1\over \sqrt{2}}A(\pi^+ K^0) = {f_K\over f_{\pi}}|{V_{us}\over
V_{ud}}|A(\pi^+\pi^0)~,\eqno(23)
$$
and a similar relation holds among the corresponding $B^-$ decay amplitudes.
These two relations are analogous to
Eqs.(22). They are descrlibed by two triangles very similar to those of
Fig. 4. In the present case the two triangles share a common base given by
$A(\pi^- \Kbar)=A(\pi^+ K^0)$, and the angle between the sides describing
$A(\pi^-\pi^0)$ and $A(\pi^+\pi^0)$ is $2\gamma$. Measurements of the four
rates
into $\pi^0 K^+, \pi^0 K^-, \pi^+ K^0, \pi^+\pi^0$, suffices to determine
$\gamma$.

This method is not as clean as the one using $B^{\pm}\to D^0K^{\pm}$ decays.
Contributions from electroweak penguin diagrams, which do not obey the above
isospin relation, were neglected. Such terms do not cancel on the
left-hand-side of (23). Although these terms are suppressed by a factor of
about
$\lambda$ (as in (18)) relative to strong penguin terms which dominate the two
amplitudes of $B^+\to\pi K$, they spoil eq.(23)
which relates the difference of these two amplitudes to the CKM-suppressed
amplitude on the right-hand-side.

One way to recover an SU(3) triangle relation, with the inclusion of
electroweak penguin terms, is to use final states which involve the octet
component of the $\eta$. One finds:
$$
A(\pi^0 K^+)+\sqrt{2}A(\pi^+ K^0)=\sqrt{3}A(\eta_8 K^+)~.\eqno(24)
$$
These amplitudes and their corresponding charge-conjugates can
be related to
$A(\pi^+\pi^0)$ as shown in Fig. 6, which could in principle determine
$\gamma$.
The problem here is that one would have to extract the amplitude into $\eta_8$
from $B$ decay measurements into  states involving $\eta$ and $\eta'$. In
addition there are corrections to (24) from SU(3) breaking terms.

Another way to resolve the uncertainty due to electroweak penguin terms is
to use, in addition to $B\to\pi K$ decays also $B_s\to\pi^0\eta_8$.
These amplitudes obey a quadrangle relation, from which $\gamma$ may
in principle be determined. The problem here is that, while $B\to\pi K$
are expected to have branching ratios of about $10^{-5}$, the branching
ratio of the electroweak penguin dominated $B_s\to\pi^0\eta_8$ decay is
estimated to be less than $10^{-6}$.

\vskip 0.5truecm
\noindent
{\it 4.4 Large final state phases from interference between resonance and
background }~[25]
\vskip 3mm

CP asymmetries in charged $B$ decays are prportional to a sine of the
final-state phase-difference of two interfering amplitudes (see (21)). So far
there exists no experimental evidence for final state phases in $B$ decays,
and it has been often assumed that such phases are likely to be small in
decays to two light high momentum particles. This would lead
to small asymmetries. Evidence for strong phases,
related to final states with well-defined isospin and angular momentum,
can be obtained from $B\to\overline{D}\pi$ decays. The amplitudes
into $D^-\pi^+,~\overline{D}^0\pi^0,~\overline{D}^0\pi^+$ obey a triangle
relation, from which the phase-difference between the $I=1/2$ and $I=3/2$
amplitudes may be determined. The present branching ratios of these decays
already imply an upper limit [26], $\delta_{1/2}-\delta_{3/2}<35^{\circ}$.
Improved measurements of these braching ratios may lead to more stringent
bounds.
Assuming, as may turn out to be the case, that final state phase differences
are small in cases of interest, one would be looking for particular
circumstances in
which these phases are enhanced. Here we wish to demonstrate one such case.

Consider the decay $B^+\to\chi_{c0}\pi^+,~\chi_{c0}\to\pi^+\pi^-$, where one is
looking for a final state with three pions, two of which have an invariant
mass around $m(\chi_{c0})=3415$ MeV. The width of this $J^P=0^+$~$c\cbar$
state,
$\Gamma(\chi_{c0})=14\pm 5$ MeV, is sufficiently large to provide a large, and
in fact maximal, CP conserving phase.
The decay amplitude into three pions, where two pions are at the resonance,
consists of two terms with different CKM phases (we neglect a small penguin
term):

\item{R=} a resonating amplitude, consisting of a product of the weak decay
amplitude of $B^+\to\chi_{c0}\pi^+$ involving a real CKM factor
$V^*_{cb}V_{cd}$  ($a_w$=real), the strong decay amplitude of
$\chi_{c0}\to\pi^+\pi^-$ ($a_s$=real), and a
Breit-Wigner term for the intermediate $\chi_{c0}$.
\item{D=} a direct decay amplitude of $B^+\to\pi^+\pi^-\pi^+$ involving a CKM
factor $V^*_{ub}V_{ud}$ with phase $\gamma$, which we write as
$(d/m_B)\exp(i\gamma)$ (d=real).

$$
R=a_wa_s{\sqrt{m\Gamma}\over s-m^2+im\Gamma}~,~~~~~~D={d\over
m_B}\exp(i\gamma)~.
\eqno(25)
$$
The total amplitude is $R+D$.

The $B^+~-~B^-$ decay rate asymmetry, integrated symmetrically around the
resonance,
is given by
$$
Asym.\approx -2{d\over a_wa_s}{\sqrt{m\Gamma}\over m_B}\sin\gamma~.\eqno(26)
$$
This is a special case of (21), in which the strong phase difference between
the resonating and direct amplitudes is maximal,
$\delta_D-\delta_R=\pi/2$. Note that all strong phases other than due to the
resonance width were neglected.
The asymmetry can be expressed in terms of the corresponding branching ratios
$$
Asym.\approx f(0)\sqrt{{B(B^+\to\pi^+\pi^-\pi^+)_{nonres.}\over
B(B^+\to\chi_{c0}\pi^+)
B(\chi_{c0}\to\pi^+\pi^-)}}{\sqrt{8\pi m\Gamma}\over m_B}\sin\gamma~.
\eqno(27)
$$
$f(0)$ is the fraction of nonresonating three pion events, where the two pions
at the
resonance mass carry zero angular momentum. (Only this part of the direct
amplitude interferes with the resonance amplitude). Model-dependent
calculations
show that $f(0)$ is of order one [27]. Using $B(\chi_{c0}\to\pi^+
\pi^-)=8\times10^{-3}$ [5], and taking reasonable estimates for the yet
unmeasured branching
ratios, $B(B^+\to\pi^+\pi^-\pi^+)_{nonres.}\sim 10^{-5},
{}~B(B^+\to\chi_{c0}\pi^+)
={\rm a~few}\times10^{-5}$, one finds an asymmetry
$$
Asym.={\cal O}(1)\sin\gamma~.\eqno(28)
$$
An observation of
such a large asymmetry requires $10^8$ or at most $10^9$ $B$ mesons.

\vskip 0.5truecm
\noindent
{\bf 5. Conclusion}
\vskip 3mm

An observation of CP violation outside the $K$ meson system is extremely
important by itself.
B decays offer a large variety of such measurements which can test the CKM
origin of CP violation.
Certain CP asymmetries in $B$ decays determine
CKM phases in manners which are free of hadronic uncertainties.
This will evidently provide a more precise determination than
available today for these fundamental parameters.
Consistency between these measurements and other (mostly CP conserving)
measurements of CKM elements would confirm the CKM mechanism of CP violation.
Inconsistencies, on the other hand, may lead the way to extensions of the
Standard Model. This is a rich field which is constantly evolving, with new
ideas being developed both on the theoretical and experimental frontiers.
We should be able to enjoy its fruits by the end of this millennium.

\vfil\eject
\noindent
{\bf Acknowledgements}
\vskip 3mm

It is a pleasure to thank David Atwood, Gad Eilam, Oscar Hern\'andez, David
London, Roberto Mendel, Jonathan Rosner, Amarjit Soni and Daniel Wyler for very
enjoyable collaborations on various topics presented here. This work was
supported in part by the United States - Israel Binational Science Foundation
and by the VPR Fund, the New York Metropolitan Research Fund.

\vskip 0.5truecm
\noindent
{\bf References}
\vskip0.3truecm

\item{1.} S. L. Glashow, {\it Nucl. Phys.} {\bf 22} (1961) 579;
S. Weinberg, {\it Phys. Rev. Lett.} {\bf 19} (1967) 1264;
A. Salam, in {\it Elementary Particle Theory}, ed. N. Svartholm (Almqvist and
Wiksell, Stockholm, 1968).
\item{2.} J. H. Christenson, J. W. Cronin, V. L. Fitch and R. Turlay, {\it
Phys. Rev. Lett.} {\bf 13} (1964) 138.
\item{3.}  N. Cabibbo, {\it Phys. Rev. Lett.} {\bf 10}
(1963) 531;  M. Kobayashi and T. Maskawa, {\it Prog. Theor.~Phys.} {\bf 49}
(1973) 652.
\item{4.} I. Bigi {\it at al.}, in {\it CP Violation}, edited by C. Jarlskog
(World Scientific, Singapore, 1989); Y. Nir and H. R. Quinn, in {\it B Decays},
Revised 2nd edition, edited by S. Stone (World Scientific, Singapore, 1994),
p. 520; M. Gronau, {\it Proceedings of Neutrino 94, XVI International
Conference on Neutrino Physics and Astrophysics}, Eilat, Israel, May 29 - June
3, 1994, edited by A. Dar, G. Eilam and M. Gronau, {\it Nucl. Phys. (Proc.
Suppl.)} {\bf B38} (1995) 136.
\item{5.} Particle Data Group, L. Montanet {\it et al., Phys. Rev.}
{\bf D50} (1994) 1173.
\item{6.} J. L. Rosner, in {\it B Decays} ({\it op. cit.}), p. 470; A. Ali
and D. London, {\it Zeit. Phys.} {\bf C65} (1995) 431.
\item{7.} A.B. Carter and A.I. Sanda, {\it Phys. Rev. Lett.} {\bf
45} (1980) 952; {\it Phys. Rev.} {\bf D23} (1981) 1567; I.I. Bigi and A.I.
Sanda,
{\it Nucl. Phys.} {\bf B193} (1981) 85; {\bf B281} (1987) 41; I. Dunietz
and J.L.
Rosner, {\it Phys. Rev.} {\bf D34} (1986) 1404.
\item{8.} CLEO Collaboration, M. S. Alam {\it et al.}, {\it Phys. Rev.} {\bf
D50} (1994) 43.
\item{9.} CLEO Collaboration, M. Battle {\it et al.}, {\it Phys. Rev. Lett.}
{\bf 71} (1993) 3922.
\item{10.} M. Gronau, {\it Phys. Lett.} {\bf B233} (1989) 479.
\item{11.} R. Aleksan, I. Dunietz, B. Kayser and F. Le Diberder,
{\it Nucl. Phys.} {\bf B361} (1991) 141; R. Aleksan, I. Dunietz and B. Kayser,
{\it Zeit. Phys.} {\bf C54} (1992) 653.
\item{12.} M. Gronau, {\it Phys. Rev. Lett.} {\bf 63} (1989) 1451;
D. London and R.D. Peccei, {\it Phys. Lett.}
{\bf B223} (1989) 257; B. Grinstein, {\it Phys. Lett.} {\bf B229} (1989) 280.
\item{13.} M. Gronau, {\it Phys. Lett.} {\bf B300} (1993) 163.
\item{14.} M. Gronau and D. London, {\it Phys. Rev. Lett.} {\bf 65} (1990)
3381.
\item{15.} N. G. Deshpande and X.-G. He, {\it Phys. Rev. Lett.} {\bf 74} (1995)
26, 4099(E).
\item{16.} M. Gronau, O. F. Hern\'andez, D. London and J. L. Rosner,
TECHNION-PH-95-11, hep-ph/9504327, to be published in {\it Phys. Rev.} {\bf D}.
\item{17.} H. J. Lipkin, Y. Nir, H. R. Quinn and A. E. Snyder, {\it Phys. Rev.}
{\bf D44} (1991) 1454; M. Gronau, {\it Phys. Lett.} {\bf B265} (1991) 389;
H. R. Quinn and A. E. Snyder, {\it Phys. Rev.} {\bf D48} (1993) 2139.
\item{18.} M. Gronau and D. Wyler, {\it Phys. Lett.} {\bf B265} (1991) 172. See
also M. Gronau and D. London, {\it Phys. Lett.} {\bf B253} (1991) 483; I.
Dunietz, {\t Phys. Lett} {\bf B270} (1991) 75.
\item{19.} S. Stone, in {\it Beauty 93, Proceedings of
the First International Workshop on $B$ Physics at Hadron Machines}, Liblice
Castle, Melnik, Czech Republic, Jan. 18-22, 1993; ed. P. E. Schlein, {\it
Nucl. Instrum. Meth.} {\bf 333} (1993) 15.
\item{20.} D. Atwood, G. Eilam, M. Gronau and A. Soni, {\it Phys. Lett.}
{\bf B341} (1995) 372.
\item{21.}  M. Gronau, J. L. Rosner and D. London, {\it
Phys. Rev. Lett.} {\bf 73} (1994) 21.
\item{22.} Deshpande and He, ref. 15 and OITS-576 (1995), hep-ph/9505369;
M. Gronau {\it et al.}, ref. 16; A. J. Buras and R. Fleischer, MPI-PhT/95-68,
hep-ph/9507303.
\item{23.} D. Zeppenfeld, {\it Zeit. Phys.} {\bf C8} (1981) 77; M. Savage and
M. Wise, {\it Phys. Rev.} {\bf D39} (1989) 3346; {\it ibid.} {\bf 40} (1989)
3127(E); J. Silva and L. Wolfenstein, {\it Phys. Rev.} {\bf D49} (1994) R1151.
\item{24.}  M. Gronau, O. F. Hern\'andez, D. London, and J. L. Rosner, {\it
Phys.  Rev.} {\bf D50} (1994) 4529; {\it Phys. Lett.} {\bf B333} (1994) 500;
and TECHNION-PH-95-10, hep-ph/9504326, to be published in {\it Phys. Rev.}
{\bf D}; L. Wolfenstein, {\it Phys. Rev.} {\bf D52} (1995) 537; A. J. Buras
and R. Fleischer, {\it Phys. Lett.} {\bf B341} (1995) 379.
\item{25.} G. Eilam, M. Gronau and R. R. Mendel, {\it Phys. Rev. Let.} {\bf 74}
(1995) 4984.
\item{26.} H. Yamamoto, HUTP-94/A006 (1994).
\item{27.} N. G. Deshpande, G. Eilam, X.-G. He and J. Trampetic, OITS-571,
hep-ph/9503273.

\vfill \eject
{}~~
\vskip0.3truecm

{\bf Figure captions}
\vskip0.3truecm

Fig. 1: The CKM unitarity triangle

Fig. 2: Isospin triangles of $B\to\pi\pi$, including electroweak
penguin corrections

Fig. 3: Diagrams describing $B^+\to\Dbar K^+$ and $B^+\to D^0 K^+$

Fig. 4: Triangles describing Eqs.(22)

Fig. 5: Penguin and tree diagrams of $B^+\to\pi^0 K^+$

Fig. 6: Triangles describing Eq.(24), its charge-conjugate and a
relation with $A(\pi^+\pi^0)$
\vfil\eject
{}~~~
\vskip10truecm
\centerline{Figure 1}
\vskip10truecm
\centerline{Figure 2}
\vfil\eject
{}~~~
\vskip10truecm
\centerline{Figure 3}
\vskip10truecm
\centerline{Figure 4}
\vfil\eject
{}~~~
\vskip10truecm
\centerline{Figure 5}
\vskip10truecm
\centerline{Figure 6}

\bye